\documentclass[preprint,authoryear,12pt]{elsarticle}
\usepackage{natbib}
\usepackage{amssymb}
\usepackage{color}
\begin{document}
\bibliographystyle{plainnat}
\begin{frontmatter}
\title{Unified Description of Matrix Mechanics and Wave Mechanics I}
\author{Yongqin Wang$^{1}$, Lifeng Kang$^{2}$}
\address{$^{1}$Department of Physics, Nanjing University, Nanjing 210008, China\\e-mail:yhwnju@hotmail.com}
\address{$^{2}$School of Pharmacy, Faculty of Medicine and Health, University of Sydney, NSW 2006, Australia}
\begin{abstract}
In this article, we discard the bra-ket notation and its correlative definitions, given by Paul Dirac. The quantum states are only described by the wave
functions. The fundamental concepts and definitions in quantum mechanics is simplified. The operator, wave functions and square matrix are represented
in the same expression which directly corresponds to the system of equations without additional introduction of the matrix representation of operator. It can
make us to convert the operator relations into the matrix relations. According to the relations between the matrices, the matrix elements will be determined.
Furthermore, the first order differential equations will be given to find the solution of equations. As a result, we unified the descriptions
of the matrix mechanics and the wave mechanics.
\end{abstract}
\begin{keyword}
  Matrix Operator Wave function Quantum mechanics
\end{keyword}
\end{frontmatter}

\section{Introduction}
In 1925, based on Niels Bohr's correspondence principle, Werner Heisenberg represented the spatial coordinate q and the momentum p by the following form
$^{[1]}$
\begin{displaymath}
q=[q(nm)e^{2\pi i\nu (nm)t}], p=[p(nm)e^{2\pi i\nu (nm)t}]
\end{displaymath}
Max Born and Pascual Jordan had discarded the above expressions in favor of the shorter notation.
\begin{displaymath}
q=q(nm), p=p(nm)
\end{displaymath}
They wrote q substituted for q(nm) as a matrix $^{[2]}$
\begin{displaymath}
\left[
\begin{array}{cccccc}
0 & q(01) & 0 & 0 & 0 & \cdots \\
q(10) & 0 & q(12) & 0 & 0 & \cdots \\
0 & q(21) & 0 & q(23) & 0 & \cdots \\
\cdots & \cdots & \cdots & \cdots & \cdots & \cdots \\
\end{array}
\right]
\end{displaymath}
In January 1926, Erwin Schr$\ddot{o}$dinger presented what is now known as the Schr$\ddot{o}$dinger equation $^{[3]}$.
\begin{displaymath}
i\hbar \frac{\partial \psi(\vec{r},t)}{\partial t}=\hat{H}\psi(\vec{r},t)
\end{displaymath}
where i is the imaginary unit, $\hbar$ is the Planck constant divided by 2$\pi$, $\psi$ is the wave function of the quantum system, $\vec{r}$ and t are the
position vector and time respectively, and $\hat{H}$ is the Hamiltonian operator. This equation formulated the wave mechanics representation of quantum
mechanics. For one-dimensional harmonic oscillator, the time-independent Schr$\ddot{o}$dinger equation is
\begin{displaymath}
(-\frac{\hbar^{2}}{2m}\frac{d^{2}}{dx^{2}}+\frac{1}{2} m \omega^{2} x^{2})\psi=E\psi
\end{displaymath}
This equation is solved and we can find
\begin{equation}
\begin{array}{cc}
\psi_{n}=\frac{H_{n}(z)}{\sqrt{2^{n-1}(n-1)!}}(\frac{m\omega}{\pi\hbar})^{\frac{1}{4}}e^{-\frac{z^{2}}{2}}
& (n=1,2,\dots)
\end{array}
\end{equation}
where the functions $H_{n}(z)$ are the Hermitian polynomials
\begin{displaymath}
\begin{array}{cc}
H_{n}(z)=(-1)^{n-1}e^{z^{2}}\frac{d^{n-1}}{dz^{n-1}}e^{-z^{2}} & (z=\sqrt{\frac{m\omega}{\hbar}}x)
\end{array}
\end{displaymath}
When the operators $\hat{a}$ and $\hat{a}^{+}$ in (16) act on the wave function $\psi_{n}$ in (1) respectively, we can obtain (17) and (18).\\
The founders of matrix mechanics tried to describe the mechanics quantum by the square matrix. They had got some success, but the origin of the square matrix
could not be explained. For the origin of matrices in matrix mechanics, we provide the following explanation.\\
In linear algebra, the product of row matrix and the same order column matrix is equal to a polynomial. For example
\begin{displaymath}
\left[
\begin{array}{cc}
A_{1}&A_{2}
\end{array}
\right] \left[
\begin{array}{c}
B_{1}\\
B_{2}
\end{array}
\right] =A_{1}B_{1}+A_{2}B_{2}
\end{displaymath}
Now that equality holds from left to right, the equality should hold from right to left
\begin{displaymath}
A_{1}B_{1}+A_{2}B_{2}= \left[
\begin{array}{cc}
A_{1} & A_{2}
\end{array}
\right] \left[
\begin{array}{c}
B_{1}\\
B_{2}
\end{array}
\right]
\end{displaymath}
So a polynomial can be expanded to the product of row matrix and the same order column matrix. When the quantum number n takes 1, 2, $\cdots$, $s\rightarrow\infty$, we can obtain the following expression from (17).
\begin{displaymath}
\hat{a}\left[
\begin{array}{ccccc}
\psi_{1} & \psi_{2} & \cdots & \psi_{s-1} &  \psi_{s} \\
\end{array}
\right]
=\left[
\begin{array}{ccccc}
\hat{a}\psi_{1} & \hat{a}\psi_{2} & \cdots & \hat{a}\psi_{s-1} &  \hat{a}\psi_{s} \\
\end{array}
\right]
\end{displaymath}
\begin{displaymath}
=\left[
\begin{array}{ccccc}
0 & \color{red}{\psi_{1}} & \cdots & \sqrt{s-2}\psi_{s-2} & \sqrt{s-1}\psi_{s-1} \\
\end{array}
\right]
\end{displaymath}
\begin{equation}
=\left[
\begin{array}{ccccc}
\color{red}{\psi_{1}} & \color{red}{\psi_{2}} & \cdots & \color{red}{\psi_{s-1}} & \color{red}{\psi_{s}} \\
\end{array}
\right]
\left[
\begin{array}{ccccc}
0 & \color{red}{1} & \cdots & 0 & 0 \\
0 & \color{red}{0} & \cdots & 0 & 0 \\
\cdots & \color{red}{\cdots} & \cdots & \cdots & \cdots \\
0 & \color{red}{0} & \cdots & 0 & \sqrt{s-1} \\
0 & \color{red}{0} & \cdots & 0 & 0 \\
\end{array}
\right]
\end{equation}
If the row matrix has a fixed array of wave functions $\psi_{1}, \psi_{2}, \cdots, \psi_{s-1}, \psi_{s}$, the square matrix is uniquely determined. The operator, wave functions and square matrix are represented in the same representation (2) which directly corresponds to the system of equations (17). As a result, the square matrix in matrix mechanics is derived from the system of equations.\\
In fact, the differential and indefinite integral are inverse operation with each other. When the operator $\hat{a}$ and the square matrix in (2) are known, (1) should be also theoretically derived from (17). Of course, given the ground state wave function $\psi_{1}$, it is easier to get other wave functions in (1) by (20).\\
In 1930, Paul Dirac published his book $\emph{Principles of Quantum Mechanics}$. In that book, the mechanical quantities were described by the operators.
Dirac established the bra-ket notations which were used to describe the quantum states and achieved great success. For one-dimensional harmonic oscillator,
Dirac obtained
\begin{displaymath}
\begin{array}{cc}
\hat{a} |N\rangle=\sqrt{N}|N-1\rangle & (N=0,1,2,\dots)
\end{array}
\end{displaymath}
\begin{displaymath}
\begin{array}{cc}
\hat{a}^{+}|N\rangle=\sqrt{N+1}|N+1\rangle & (N=0,1,2,\dots)
\end{array}
\end{displaymath}
Obviously, these two expressions are similar to (17) and (18) respectively. The ladder operator method, developed by Dirac, allows us to extract the eigenvalues
without directly solving the differential equation. However, Dirac had to introduce the additional bra-ket notation and its correlative definitions. For one-dimensional harmonic oscillator,
\begin{displaymath}
\begin{array}{cc}
\langle N|\hat{a}^{+}\hat{a}|N\rangle=(\hat{a}|N\rangle)^{\dag}\hat{a}|N\rangle=|\hat{a}|N\rangle|^{2}\geq 0
\end{array}
\end{displaymath}
where $\dag$ denotes conjugate transpose. In order to make the above expression hold, Dirac had to give the following definitions\\
1, Not only the operator $\hat{a}$ acts on $|N\rangle$, but also $\hat{a}^{+}$ acts on $\langle N|$.\\
2, $\langle N|\hat{a}^{+}=(\hat{a}|N\rangle)^{\dag}$\\
However, these definitions are not necessary. If the ket notation is replaced by the wave function, the bra-ket notation and its correlative definitions by Dirac become redundant with invoking assistance from the matrix tool. It is necessary for us to reconstruct the mathematical foundations of quantum mechanics so that the fundamental concepts and definitions in quantum mechanics can be simplified.
\section{Mathematical foundations}
In quantum mechanics, the wave function describes quantum state. Therefore, two definitions on wave function are given as follows\\
\textbf{Definition 1} If $\phi=\sum\limits_{n=1}^{s}c_{n}\psi_{n}$, then $|\phi\rangle=\sum\limits_{n=1}^{s}c_{n}|\psi_{n}\rangle$ and
$\langle\phi|=\sum\limits_{n=1}^{s}c_{n}^*\langle\psi_{n}|\,$ respectively,
where * denotes complex conjugation.\\
\textbf{Definition 2} If $\phi$ and $\varphi$ are two arbitrary wave functions, then the scalar product of $\phi$ and $\varphi$ is
$\langle\phi|\varphi\rangle=\int{\phi^*\varphi}d\tau$.\\
where the symbols $|\rangle$ and $\langle|$ which meanings are only limited to the above definitions can not be used to denote abstract vectors and linear
functionals in mathematics. \textcolor{blue}{We will discard the following concepts and definitions, given by Diarc:\\
	1, The quantum states were described by the ket notation $|\alpha\rangle$.\\
	2, The relation between the bra and ket notations $\langle\alpha|=|\alpha\rangle^{\dag}$.\\
	3, The operator $\hat{A}$ operates on the ket $|\alpha\rangle$ and the bra $\langle\alpha|$.\\
	4, The matrix representation of the operator.\\
	5, $|\alpha\rangle\langle\beta|$ appears as a linear operator.}\\
In quantum mechanics, the mechanical quantities are described by the operators which are Hermitian operators. The operator acts not on the symbols $|\rangle$ and $\langle|$, but on the functions in the new system. The definition on the operator is given as follows\\
\textbf{Definition 3} If $\hat{F}$ is a Hermitian operator, then for two arbitrary wave functions
$\phi$ and $\varphi$, we have
\begin{displaymath}
\langle\phi|\hat{F}\varphi\rangle=\langle\varphi|\hat{F}\phi\rangle^{*}
\end{displaymath}
According to Born's probability interpretation and matrix transformation, we can always assume that $\psi_{1},\psi_{2},\cdots,\psi_{s}$ are a set of
orthonormalized wave functions. Simultaneously
\begin{displaymath}
\begin{array}{cc}
\hat{D}\psi_{n}=D_{n}\psi_{n} & (n=1,2,\cdots,s)
\end{array}
\end{displaymath}
where $D_{1},D_{2},\cdots,D_{s}$ are called eigenvalues of the operator $\hat{D}$ corresponding to the eigenfunctions.\\
When an operator $\hat{F}$ acts on the orthonormalized wave functions $\psi_{1},\psi_{2},\cdots,\psi_{s}$, we will obtain some new wave functions
$\phi_{1}=\hat{F}\psi_{1},\phi_{2}=\hat{F}\psi_{2},\cdots,\phi_{s}=\hat{F}\psi_{s}$. According to the principle of superposition states in
quantum mechanics, these new wave functions $\phi_{1},\phi_{2},\cdots,\phi_{s}$ were represented by linear combination of orthonormalized wave
functions $\psi_{1},\psi_{2}, \cdots,\psi_{s}$.
\begin{small}
	\begin{equation}
	\hat{F}\psi_{1}=\phi_{1}=\sum\limits_{n=1}^{s}F_{n1}\psi_{n},\hat{F}\psi_{2}=\phi_{2}=\sum\limits_{n=1}^{s}F_{n2}\psi_{n},\cdots,
	\hat{F}\psi_{s}=\phi_{s}=\sum\limits_{n=1}^{s}F_{ns}\psi_{n}
	\end{equation}
\end{small}
According to definition 1 and the inverse law of matrix multiplication, similar to (2), (3) can be written as
\begin{displaymath}
\left[\setlength\arraycolsep{0.2em}
\begin{array}{cccc}
|\hat{F}\psi_{1}\rangle & |\hat{F}\psi_{2}\rangle & \cdots & |\hat{F}\psi_{s}\rangle \\
\end{array}
\right]
=
\left[\setlength\arraycolsep{0.2em}
\begin{array}{cccc}
|\sum\limits_{n=1}^{s}F_{n1}\psi_{n}\rangle & |\sum\limits_{n=1}^{s}F_{n2}\psi_{n}\rangle & \cdots & |\sum\limits_{n=1}^{s}F_{ns}\psi_{n}\rangle \\
\end{array}
\right]
\end{displaymath}
\begin{displaymath}
=
\left[\setlength\arraycolsep{0.2em}
\begin{array}{cccc}
\sum\limits_{n=1}^{s}F_{n1}|\psi_{n}\rangle & \sum\limits_{n=1}^{s}F_{n2}|\psi_{n}\rangle & \cdots & \sum\limits_{n=1}^{s}F_{ns}|\psi_{n}\rangle \\
\end{array}
\right]
\end{displaymath}
\begin{displaymath}
=
\left[\setlength\arraycolsep{0.2em}
\begin{array}{cccc}
|\psi_{1}\rangle & |\psi_{2}\rangle & \cdots & |\psi_{s}\rangle \\
\end{array}
\right]
\left[
\begin{array}{cccc}
F_{11} & F_{12} & \cdots & F_{1s} \\
F_{21} & F_{22} & \cdots & F_{2s} \\
\cdots & \cdots & \cdots & \cdots \\
F_{s1} & F_{s2} & \cdots & F_{ss} \\
\end{array}
\right]
\end{displaymath}
Both sides of the above expression are left multiplied by the column matrix
$\left[
\begin{array}{cccc}
\langle\psi_{1}| & \langle\psi_{2}| & \cdots & \langle\psi_{s}| \\
\end{array}
\right]^{T}$. Using the orthonormality of wave functions $\psi_{1},\psi_{2},\cdots,\psi_{s}$,
\begin{displaymath}
\left[
\begin{array}{cccc}
\langle\psi_{1}|\hat{F}\psi_{1}\rangle & \langle\psi_{1}|\hat{F}\psi_{2}\rangle & \cdots & \langle\psi_{1}|\hat{F}\psi_{s}\rangle \\
\langle\psi_{2}|\hat{F}\psi_{1}\rangle & \langle\psi_{2}|\hat{F}\psi_{2}\rangle & \cdots & \langle\psi_{2}|\hat{F}\psi_{s}\rangle \\
\cdots & \cdots & \cdots & \cdots \\
\langle\psi_{s}|\hat{F}\psi_{1}\rangle & \langle\psi_{s}|\hat{F}\psi_{2}\rangle & \cdots & \langle\psi_{s}|\hat{F}\psi_{s}\rangle \\
\end{array}
\right]
=
\left[
\begin{array}{cccc}
F_{11} & F_{12} & \cdots & F_{1s} \\
F_{21} & F_{22} & \cdots & F_{2s} \\
\cdots & \cdots & \cdots & \cdots \\
F_{s1} & F_{s2} & \cdots & F_{ss} \\
\end{array}
\right]
\end{displaymath}
The square matrix on the right side of the above equation is the same as the one in the following expression.
\begin{displaymath}
\hat{F}\left[
\begin{array}{cccc}
\psi_{1} & \psi_{2} & \cdots & \psi_{s} \\
\end{array}
\right]
=
\left[
\begin{array}{cccc}
\psi_{1} & \psi_{2} & \cdots & \psi_{s} \\
\end{array}
\right]
\left[
\begin{array}{cccc}
F_{11} & F_{12} & \cdots & F_{1s} \\
F_{21} & F_{22} & \cdots & F_{2s} \\
\cdots & \cdots & \cdots & \cdots \\
F_{s1} & F_{s2} & \cdots & F_{ss} \\
\end{array}
\right]
\end{displaymath}
Thus we don't need to introduce the concepts on the matrix representation of operator. Furthermore, if $\hat{F}$ is a Hermitian operator, then
\begin{displaymath}
\begin{array}{cc}
F_{ij}=\langle\psi_{i}|\hat{F}\psi_{j}\rangle=\langle\psi_{j}|\hat{F}\psi_{i}\rangle^{*}=F_{ji}^{*} & (i=1,2,\cdots,s; j=1,2,\cdots,s)
\end{array}
\end{displaymath}
Therefore\\
\begin{equation}
\hat{F}\left[
\begin{array}{cccc}
\psi_{1} & \psi_{2} & \cdots & \psi_{s} \\
\end{array}
\right]
=
\left[
\begin{array}{cccc}
\psi_{1} & \psi_{2} & \cdots & \psi_{s} \\
\end{array}
\right]
\left[
\begin{array}{cccc}
F_{11} & F_{21}^{*} & \cdots & F_{s1}^{*} \\
F_{21} & F_{22} & \cdots & F_{s2}^{*} \\
\cdots & \cdots & \cdots & \cdots \\
F_{s1} & F_{s2} & \cdots & F_{ss} \\
\end{array}
\right]
\end{equation}
where $F_{11}, F_{22}, \cdots, F_{ss}$ are real numbers.\\
\textbf{Theorem}: Let $\psi_{1},\psi_{2},\cdots,\psi_{s}$ be a set of orthonormalized wave functions.
\begin{equation}
\begin{array}{cc}
\hat{D}\psi_{n}=D_{n}\psi_{n} & (n=1,2,\cdots,s)
\end{array}
\end{equation}
where $D_{1},D_{2},\cdots,D_{s}$ are called eigenvalues of the operator $\hat{D}$ corresponding to the eigenfunctions and $D_{1}<D_{2}<\cdots<D_{s}$.
The operators $\hat{D}$, $\hat{F}$ and $\hat{G}$ are the Hermitian operators and obey the following canonical commutation relations
\begin{equation}
\hat{G}=ic_{1}[\hat{D},\hat{F}],\hat{F}=ic_{2}[\hat{G},\hat{D}]
\end{equation}
If (4) holds, then
\begin{equation}
\begin{array}{cc}
D_{n}=D_{1}+\frac{n-1}{\sqrt{c_{1}c_{2}}} & (n=1,2,\cdots,s)
\end{array}
\end{equation}
Let $\hat{f}=\hat{F}+i\frac{\sqrt{c_{1}c_{2}}}{c_{1}}\hat{G}$ and $\hat{g}=\hat{F}-i\frac{\sqrt{c_{1}c_{2}}}{c_{1}}\hat{G}$, we can obtain
\begin{displaymath}
\hat{f}\left[
\begin{array}{ccccc}
\psi_{1} & \psi_{2} & \cdots & \psi_{s-1} & \psi_{s} \\
\end{array}
\right]
=
\left[
\begin{array}{ccccc}
\psi_{1} & \psi_{2} & \cdots & \psi_{s-1} & \psi_{s} \\
\end{array}
\right]
\end{displaymath}
\begin{equation}
2\left[
\begin{array}{ccccc}
0 & F_{21}^{*} & \cdots & 0 & 0 \\
0 & 0 & \cdots & 0 & 0 \\
\cdots & \cdots & \cdots & \cdots \\
0 & 0 & \cdots & 0 & F_{ss-1}^{*} \\
0 & 0 & \cdots & 0 & 0 \\
\end{array}
\right]
\end{equation}
\begin{displaymath}
\hat{g}\left[
\begin{array}{ccccc}
\psi_{1} & \psi_{2} & \cdots & \psi_{s-1} & \psi_{s} \\
\end{array}
\right]
=
\left[
\begin{array}{ccccc}
\psi_{1} & \psi_{2} & \cdots & \psi_{s-1} & \psi_{s} \\
\end{array}
\right]
\end{displaymath}
\begin{equation}
2\left[
\begin{array}{ccccc}
0 & 0 & \cdots & 0 & 0 \\
F_{21} & 0 & \cdots & 0 & 0 \\
\cdots & \cdots & \cdots & \cdots \\
0 & 0 & \cdots & 0 & 0 \\
0 & 0 & \cdots & F_{ss-1} & 0 \\
\end{array}
\right]
\end{equation}
\emph{Proof}
Let $\psi_{1},\psi_{2},\cdots,\psi_{s}$ be a set of orthonormalized wave functions.
\begin{displaymath}
\begin{array}{cc}
\hat{D}\psi_{n}=D_{n}\psi_{n} & (n=1,2,\cdots,s)
\end{array}
\end{displaymath}
where $D_{1},D_{2},\cdots,D_{s}$ are called eigenvalues of the operator $\hat{D}$ corresponding to the eigenfunctions.
\begin{displaymath}
\hat{D}\left[
\begin{array}{cccc}
\psi_{1} & \psi_{2} & \cdots & \psi_{s} \\
\end{array}
\right]
=
\left[
\begin{array}{cccc}
\psi_{1} & \psi_{2} & \cdots & \psi_{s} \\
\end{array}
\right]
\left[
\begin{array}{cccc}
D_{1} & 0 & \cdots & 0 \\
0 & D_{2} & \cdots & 0 \\
\cdots & \cdots & \cdots & \cdots \\
0 & 0 & \cdots & D_{s} \\
\end{array}
\right]
\end{displaymath}\\
Because $\hat{F}$ is a Hermitian operator, it is assumed that (4) holds. The operator, wave functions and square matrix are represented in the same expression. It can make us to convert the operator relations into the square matrix
relations.
\begin{displaymath}
\hat{G}\left[
\begin{array}{cccc}
\psi_{1} & \psi_{2} & \cdots & \psi_{s} \\
\end{array}
\right]
=ic_{1}[\hat{D},\hat{F}]
\left[
\begin{array}{cccc}
\psi_{1} & \psi_{2} & \cdots & \psi_{s} \\
\end{array}
\right]
\end{displaymath}
\begin{displaymath}
=ic_{1}\hat{D}\color{red}{\hat{F}
	\left[
	\begin{array}{cccc}
	\psi_{1} & \psi_{2} & \cdots & \psi_{s} \\
	\end{array}
	\right]}
\color{black}{-ic_{1}\hat{F}}\hat{D}
\left[
\begin{array}{cccc}
\psi_{1} & \psi_{2} & \cdots & \psi_{s} \\
\end{array}
\right]
\end{displaymath}
\begin{displaymath}
=ic_{1}\hat{D}
\color{red}{\left[
	\begin{array}{cccc}
	\psi_{1} & \psi_{2} & \cdots & \psi_{s} \\
	\end{array}
	\right]
	\left[
	\begin{array}{cccc}
	F_{11} & F_{21}^{*} & \cdots & F_{s1}^{*} \\
	F_{21} & F_{22} & \cdots & F_{s2}^{*} \\
	\cdots & \cdots & \cdots & \cdots \\
	F_{s1} & F_{s2} & \cdots & F_{ss} \\
	\end{array}
	\right]}
\end{displaymath}
\begin{displaymath}
-ic_{1}\color{red}{\hat{F}
	\left[
	\begin{array}{cccc}
	\psi_{1} & \psi_{2} & \cdots & \psi_{s} \\
	\end{array}
	\right]}
\color{black}{\left[
	\begin{array}{cccc}
	D_{1} & 0 & \cdots & 0 \\
	0 & D_{2} & \cdots & 0 \\
	\cdots & \cdots & \cdots & \cdots \\
	0 & 0 & \cdots & D_{s} \\
	\end{array}
	\right]}
\end{displaymath}
\begin{displaymath}
=ic_{1}
\left[
\begin{array}{cccc}
\psi_{1} & \psi_{2} & \cdots & \psi_{s} \\
\end{array}
\right]
\left[
\begin{array}{cccc}
D_{1} & 0 & \cdots & 0 \\
0 & D_{2} & \cdots & 0 \\
\cdots & \cdots & \cdots & \cdots \\
0 & 0 & \cdots & D_{s} \\
\end{array}
\right]
\left[
\begin{array}{cccc}
F_{11} & F_{21}^{*} & \cdots & F_{s1}^{*} \\
F_{21} & F_{22} & \cdots & F_{s2}^{*} \\
\cdots & \cdots & \cdots & \cdots \\
F_{s1} & F_{s2} & \cdots & F_{ss} \\
\end{array}
\right]
\end{displaymath}
\begin{displaymath}
-ic_{1}
\color{red}{\left[
	\begin{array}{cccc}
	\psi_{1} & \psi_{2} & \cdots & \psi_{s} \\
	\end{array}
	\right]
	\left[
	\begin{array}{cccc}
	F_{11} & F_{21}^{*} & \cdots & F_{s1}^{*} \\
	F_{21} & F_{22} & \cdots & F_{s2}^{*} \\
	\cdots & \cdots & \cdots & \cdots \\
	F_{s1} & F_{s2} & \cdots & F_{ss} \\
	\end{array}
	\right]}
\color{black}{\left[
	\begin{array}{cccc}
	D_{1} & 0 & \cdots & 0 \\
	0 & D_{2} & \cdots & 0 \\
	\cdots & \cdots & \cdots & \cdots \\
	0 & 0 & \cdots & D_{s} \\
	\end{array}
	\right]}
\end{displaymath}
\begin{displaymath}
=\!\left[\setlength\arraycolsep{0.1em}
\begin{array}{cccc}
\psi_{1} & \psi_{2} & \cdots & \psi_{s} \\
\end{array}
\right]\!
ic_{1}(\left[\setlength\arraycolsep{0.1em}
\begin{array}{cccc}
D_{1}F_{11} & D_{1}F_{21}^{*} & \cdots & D_{1}F_{s1}^{*} \\
D_{2}F_{21} & D_{2}F_{22} & \cdots & D_{2}F_{s2}^{*} \\
\cdots & \cdots & \cdots & \cdots \\
D_{s}F_{s1} & D_{s}F_{s2} & \cdots & D_{s}F_{ss} \\
\end{array}
\right]
\!-\!\left[\setlength\arraycolsep{0.2em}
\begin{array}{cccc}
D_{1}F_{11} & D_{2}F_{21}^{*} & \cdots & D_{s}F_{s1}^{*} \\
D_{1}F_{21} & D_{2}F_{22} & \cdots & D_{s}F_{s2}^{*} \\
\cdots & \cdots & \cdots & \cdots \\
D_{1}F_{s1} & D_{2}F_{s2} & \cdots & D_{s}F_{ss} \\
\end{array}
\right])
\end{displaymath}
\begin{displaymath}
=\left[\setlength\arraycolsep{0.2em}
\begin{array}{cccc}
\psi_{1} & \psi_{2} & \cdots & \psi_{s} \\
\end{array}
\right]
ic_{1}\left[\setlength\arraycolsep{0.2em}
\begin{array}{cccc}
0 & (D_{1}-D_{2})F_{21}^{*} & \cdots & (D_{1}-D_{s})F_{s1}^{*} \\
(D_{2}-D_{1})F_{21} & 0 & \cdots & (D_{2}-D_{s})F_{s2}^{*} \\
\cdots & \cdots & \cdots & \cdots \\
(D_{s}-D_{1})F_{s1} & (D_{s}-D_{2})F_{s2} & \cdots & 0 \\
\end{array}
\right]
\end{displaymath}
Similarly,
\begin{displaymath}
\hat{F}\left[
\begin{array}{cccc}
\psi_{1} & \psi_{2} & \cdots & \psi_{s} \\
\end{array}
\right]
=ic_{2}[\hat{G},\hat{D}]
\left[
\begin{array}{cccc}
\psi_{1} & \psi_{2} & \cdots & \psi_{s} \\
\end{array}
\right]
\end{displaymath}
\begin{displaymath}
=\left[\setlength\arraycolsep{0.2em}
\begin{array}{cccc}
\psi_{1} & \psi_{2} & \cdots & \psi_{s} \\
\end{array}
\right]
c_{1}c_{2}\left[\setlength\arraycolsep{0.2em}
\begin{array}{cccc}
0 & (D_{2}-D_{1})^{2}F_{21}^{*} & \cdots & (D_{s}-D_{1})^{2}F_{s1}^{*} \\
(D_{2}-D_{1})^{2}F_{21} & 0 & \cdots & (D_{s}-D_{2})^{2}F_{s2}^{*} \\
\cdots & \cdots & \cdots & \cdots \\
(D_{s}-D_{1})^{2}F_{s1} & (D_{s}-D_{2})^{2}F_{s2} & \cdots & 0 \\
\end{array}
\right]
\end{displaymath}
Compared with (4),
\begin{displaymath}
\left[\setlength\arraycolsep{0.1em}
\begin{array}{cccc}
F_{11} & F_{21}^{*} & \cdots & F_{s1}^{*} \\
F_{21} & F_{22} & \cdots & F_{s2}^{*} \\
\cdots & \cdots & \cdots & \cdots \\
F_{s1} & F_{s2} & \cdots & F_{ss} \\
\end{array}
\right]
=c_{1}c_{2}\left[\setlength\arraycolsep{0.1em}
\begin{array}{cccc}
0 & (D_{2}-D_{1})^{2}F_{21}^{*} & \cdots & (D_{s}-D_{1})^{2}F_{s1}^{*} \\
(D_{2}-D_{1})^{2}F_{21} & 0 & \cdots & (D_{s}-D_{2})^{2}F_{s2}^{*} \\
\cdots & \cdots & \cdots & \cdots \\
(D_{s}-D_{1})^{2}F_{s1} & (D_{s}-D_{2})^{2}F_{s2} & \cdots & 0 \\
\end{array}
\right]
\end{displaymath}
According to the relations between the square matrices, the matrix elements will be determined without invoking assistance from the bra-ket notation and
its correlative definitions by Dirac. It is clear that
\begin{displaymath}
F_{11}=F_{22}=\cdots=F_{ss}=0
\end{displaymath}
Because $D_{1},D_{2},\cdots,D_{s}$ are real numbers and $D_{1}<D_{2}<\cdots<D_{s}$, we have
\begin{displaymath}
F_{21}=c_{1}c_{2}(D_{2}-D_{1})^{2}F_{21}\Rightarrow D_{2}-D_{1}=\frac{1}{\sqrt{c_{1}c_{2}}}
\end{displaymath}
\begin{displaymath}
F_{31}\!=\!c_{1}c_{2}(D_{3}-D_{1})^{2}F_{31},F_{32}\!=\!c_{1}c_{2}(D_{3}-D_{2})^{2}F_{32}\Rightarrow D_{3}-D_{2}\!=\!\frac{1}{\sqrt{c_{1}c_{2}}}, F_{31}=0
\end{displaymath}
\begin{displaymath}
\cdots,\cdots,
\end{displaymath}
\begin{displaymath}
F_{\!s1}\!=\!c_{1}c_{2}(D_{\!s}-D_{1})^{2}F_{\!s1},F_{\!s2}\!=\!c_{1}c_{2}(D_{\!s}-D_{2})^{2}F_{\!s2},\cdots,
F_{\!ss-\!1}\!=\!c_{1}c_{2}(D_{\!s}-D_{\!s-\!1})^{2}F_{\!ss-\!1}
\end{displaymath}
\begin{displaymath}
\Rightarrow D_{s}-D_{s-1}=\frac{1}{\sqrt{c_{1}c_{2}}}, F_{s1}=F_{s2}=\cdots=F_{ss-2}=0
\end{displaymath}
Therefore
\begin{displaymath}
\begin{array}{cc}
D_{n}=D_{1}+\frac{n-1}{\sqrt{c_{1}c_{2}}} & (n=1,2,\cdots,s)
\end{array}
\end{displaymath}
\begin{displaymath}
\hat{F}\left[
\begin{array}{ccccc}
\psi_{1} & \psi_{2} & \cdots & \psi_{s-1} & \psi_{s} \\
\end{array}
\right]
=
\left[
\begin{array}{ccccc}
\psi_{1} & \psi_{2} & \cdots & \psi_{s-1} & \psi_{s} \\
\end{array}
\right]
\end{displaymath}
\begin{displaymath}
\left[
\begin{array}{ccccc}
0 & F_{21}^{*} & \cdots & 0 & 0 \\
F_{21} & 0 & \cdots & 0 & 0 \\
\cdots & \cdots & \cdots & \cdots \\
0 & 0 & \cdots & 0 & F_{ss-1}^{*} \\
0 & 0 & \cdots & F_{ss-1} & 0 \\
\end{array}
\right]
\end{displaymath}
\begin{displaymath}
\hat{G}\left[
\begin{array}{ccccc}
\psi_{1} & \psi_{2} & \cdots & \psi_{s-1} & \psi_{s} \\
\end{array}
\right]
=
\left[
\begin{array}{ccccc}
\psi_{1} & \psi_{2} & \cdots & \psi_{s-1} & \psi_{s} \\
\end{array}
\right]
\end{displaymath}
\begin{displaymath}
\frac{ic_{1}}{\sqrt{c_{1}c_{2}}}\left[
\begin{array}{ccccc}
0 & -F_{21}^{*} & \cdots & 0 & 0 \\
F_{21} & 0 & \cdots & 0 & 0 \\
\cdots & \cdots & \cdots & \cdots \\
0 & 0 & \cdots & 0 & -F_{ss-1}^{*} \\
0 & 0 & \cdots & F_{ss-1} & 0 \\
\end{array}
\right]
\end{displaymath}
It follows by the above two expressions that we will get (8) and (9). Let us illustrate our method by taking the one-dimensional harmonic oscillator, the angular momentum and the hydrogen atom of a solvable system as examples.
\section{\bf One-dimensional Harmonic Oscillator}
The Hamiltonian of the particle is
\begin{displaymath}
\hat{H}=\frac{\hat{p}^{2}}{2m}+\frac{1}{2}m\omega^{2}\hat{x}^{2}
\end{displaymath}
where m is the particle's mass, $\omega$ is the angular frequency of the oscillator, $\hat{x}$ is the position operator and $\hat{p}$ is the momentum operator,
given respectively by
\begin{displaymath}
\begin{array}{cc}
\hat{x}=x & \hat{p}=-i\hbar\frac{d}{dx}
\end{array}
\end{displaymath}
It is easy for us to show that
\begin{displaymath}
\hat{p}=\frac{im}{\hbar}[\hat{H},\hat{x}]
\end{displaymath}
\begin{displaymath}
\hat{x}=\frac{i}{m\hbar\omega^{2}}[\hat{p},\hat{H}]
\end{displaymath}
Let $E_{1}, E_{2}, \cdots, E_{s}$ be the eigenvalues of $\hat{H}$ and $\psi_{1}, \psi_{2}, \cdots, \psi_{s}$ be the orthonormalized eigenfunctions belonging to it,
then
\begin{displaymath}
\begin{array}{cc}
\hat{H}\psi_{n}=E_{n}\psi_{n} & (n=1,2,\cdots,s)
\end{array}
\end{displaymath}
Because $\hat{x}$ is a Hermitian operator, it is assumed that
\begin{displaymath}
\hat{x}\left[
\begin{array}{cccc}
\psi_{1} & \psi_{2} & \cdots & \psi_{s} \\
\end{array}
\right]
=
\left[
\begin{array}{cccc}
\psi_{1} & \psi_{2} & \cdots & \psi_{s} \\
\end{array}
\right]
\left[
\begin{array}{cccc}
X_{11} & X_{21}^{*} & \cdots & X_{s1}^{*} \\
X_{21} & X_{22} & \cdots & X_{s2}^{*} \\
\cdots & \cdots & \cdots & \cdots \\
X_{s1} & X_{s2} & \cdots & X_{ss} \\
\end{array}
\right]
\end{displaymath}
where $X_{11}, X_{22}, \cdots, X_{ss}$ are real numbers. Let $\hat{f}=\hat{x}+\frac{i}{m\omega}\hat{p}$, $\hat{g}=\hat{x}-\frac{i}{m\omega}\hat{p}$ $(c_{1}=\frac{m}{\hbar}, c_{2}=\frac{1}{m\hbar\omega^2})$ in terms of the theorem,
\begin{equation}
\begin{array}{cc}
E_{n}=E_{1}+(n-1)\hbar \omega & (n=1,2,\cdots,s)
\end{array}
\end{equation}
\begin{displaymath}
\hat{f}\left[
\begin{array}{ccccc}
\psi_{1} & \psi_{2} & \cdots & \psi_{s-1} & \psi_{s} \\
\end{array}
\right]
=
\left[
\begin{array}{ccccc}
\psi_{1} & \psi_{2} & \cdots & \psi_{s-1} & \psi_{s} \\
\end{array}
\right]
\end{displaymath}
\begin{equation}
2\left[
\begin{array}{ccccc}
0 & X_{21}^{*} & \cdots & 0 & 0 \\
0 & 0 & \cdots & 0 & 0 \\
\cdots & \cdots & \cdots & \cdots \\
0 & 0 & \cdots & 0 & X_{ss-1}^{*} \\
0 & 0 & \cdots & 0 & 0 \\
\end{array}
\right]
\end{equation}
\begin{displaymath}
\hat{g}\left[
\begin{array}{ccccc}
\psi_{1} & \psi_{2} & \cdots & \psi_{s-1} & \psi_{s} \\
\end{array}
\right]
=
\left[
\begin{array}{ccccc}
\psi_{1} & \psi_{2} & \cdots & \psi_{s-1} & \psi_{s} \\
\end{array}
\right]
\end{displaymath}
\begin{equation}
2\left[
\begin{array}{ccccc}
0 & 0 & \cdots & 0 & 0 \\
X_{21} & 0 & \cdots & 0 & 0 \\
\cdots & \cdots & \cdots & \cdots \\
0 & 0 & \cdots & 0 & 0 \\
0 & 0 & \cdots & X_{ss-1} & 0 \\
\end{array}
\right]
\end{equation}
Since $[\hat{x},\hat{p}]=i\hbar$,
\begin{displaymath}
[\hat{f},\hat{g}]=\frac{2\hbar}{m\omega}
\end{displaymath}
Combining with (11)-(12),
\begin{displaymath}
\frac{2\hbar}{m\omega} \left[
\begin{array}{ccccc}
\psi_{1} & \psi_{2} & \cdots & \psi_{s-1} & \psi_{s} \\
\end{array}
\right]
=
[\hat{f},\hat{g}] \left[
\begin{array}{ccccc}
\psi_{1} & \psi_{2} & \cdots & \psi_{s-1} & \psi_{s} \\
\end{array}
\right]
\end{displaymath}
\begin{displaymath}
\!=\!4\! \left[\setlength\arraycolsep{0.2em}
\begin{array}{ccccc}
|X_{\!21}|^2 \psi_{1} & (|X_{\!32}|^2\!-\!|X_{\!21}|^2)\psi_{2} & \cdots & (|X_{\!ss-1}|^2\!-\!|X_{\!s-\!1\!s-\!2}|^2)\psi_{s-1} & -\!|X_{\!ss-1}|^2 \psi_{s} \\
\end{array}
\right]
\end{displaymath}
So
\begin{equation}
\begin{array}{cccc}
|X_{21}|^{2}=\frac{\hbar}{2m\omega}, & |X_{32}|^{2}=\frac{2\hbar}{2m\omega}, & \cdots, & |X_{ss-1}|^{2}=\frac{(s-1)\hbar}{2m\omega}
\end{array}
\end{equation}
and
\begin{displaymath}
-|X_{ss-1}|^{2}\psi_{s}=\frac{\hbar}{2m\omega}\psi_{s}
\end{displaymath}
Thus
\begin{equation}
\psi_{s}=0
\end{equation}
We can express $\hat{x}$ and $\hat{p}$ in terms of $\hat{f}$ and $\hat{g}$. So
\begin{displaymath}
\hat{H}=\frac{\hat{p}^2}{2m}+\frac{1}{2} m \omega^2 \hat{x}^2=\frac{1}{4} m \omega^2 (\hat{f}\hat{g}+\hat{g}\hat{f})
\end{displaymath}
Combining with (11)-(13),
\begin{displaymath}
E_{\!1}\psi_{\!1}=\hat{H}\psi_{\!1}=\frac{m}{4} \omega^{2}\hat{f}\hat{g}\psi_{\!1}+\frac{m}{4} \omega^{2}\hat{g}\hat{f}\psi_{\!1}=\frac{m}{2}\omega^{2}X_{\!21}\hat{f}\psi_{2}
=m\omega^2|X_{\!21}|^2\psi_{\!1}=\frac{1}{2}\hbar\omega \psi_{\!1}
\end{displaymath}
Thus
\begin{displaymath}
E_{1}=\frac{1}{2}\hbar\omega
\end{displaymath}
Combining with (10),
\begin{equation}
\begin{array}{cc}
E_{n}=(n-\frac{1}{2})\hbar\omega & (n=1,2,\cdots,s-1, s \rightarrow \infty)
\end{array}
\end{equation}
If we take positive real solutions from (13), then we have
\begin{displaymath}
\begin{array}{cccc}
X_{21}=\sqrt{\frac{\hbar}{2m\omega}}, & X_{32}=\sqrt{\frac{2\hbar}{2m\omega}}, & \cdots, & X_{ss-1}=\sqrt{\frac{(s-1)\hbar}{2m\omega}}
\end{array}
\end{displaymath}
From (11) and (12),
\begin{displaymath}
\begin{array}{cccc}
\hat{f}\psi_{1}=0, & \hat{f}\psi_{2}=\sqrt{\frac{2\hbar}{m\omega}}\psi_{1}, & \cdots, & \hat{f}\psi_{s}=\sqrt{\frac{2\hbar}{m\omega}}\sqrt{s-1}\psi_{s-1}
\end{array}
\end{displaymath}
\begin{displaymath}
\setlength\arraycolsep{0.3em}
\begin{array}{ccccc}
\hat{g}\psi_{1}=\sqrt{\frac{2\hbar}{m\omega}}\psi_{2}, & \hat{g}\psi_{2}=\sqrt{\frac{2\hbar}{m\omega}}\sqrt{2}\psi_{3}, & \cdots, & \hat{g}\psi_{s-1}=\sqrt{\frac{2\hbar}{m\omega}}\sqrt{s-1}\psi_{s}, & \hat{g}\psi_{s}=0
\end{array}
\end{displaymath}
Set
\begin{equation}
\begin{array}{cc}
\hat{a}=\sqrt{\frac{m\omega}{2\hbar}}\hat{f}=\sqrt{\frac{m\omega}{2\hbar}}(x+\frac{\hbar}{m\omega}\frac{d}{dx}), \hat{a}^{+}=\sqrt{\frac{m\omega}{2\hbar}}\hat{g}=\sqrt{\frac{m\omega}{2\hbar}}(x-\frac{\hbar}{m\omega}\frac{d}{dx})
\end{array}
\end{equation}
So
\begin{displaymath}
\begin{array}{cccc}
\hat{a}\psi_{1}=0, & \hat{a}\psi_{2}=\psi_{1}, & \cdots, & \hat{a}\psi_{s}=\sqrt{s-1}\psi_{s-1}
\end{array}
\end{displaymath}
\begin{displaymath}
\begin{array}{ccccc}
\hat{a}^{+}\psi_{1}=\psi_{2}, & \hat{a}^{+}\psi_{2}=\sqrt{2}\psi_{3}, & \cdots, & \hat{a}^{+}\psi_{s-1}=\sqrt{s-1}\psi_{s}, & \hat{a}^{+}\psi_{s}=0
\end{array}
\end{displaymath}
That is
\begin{equation}
\begin{array}{cc}
\hat{a} \psi_{n}=\sqrt{n-1}\psi_{n-1} & (n=1,2,\dots,s)
\end{array}
\end{equation}
\begin{equation}
\begin{array}{cc}
\hat{a}^{+}\psi_{n}=\sqrt{n}\psi_{n+1} & (n=1,2,\dots, s-1)
\end{array}
\end{equation}
and
\begin{displaymath}
\hat{a}^{+}\psi_{s}=0
\end{displaymath}
From (16) and (17),
\begin{displaymath}
\hat{a}\psi_{1}=0 \Rightarrow x \psi_{1}+\frac{\hbar}{m \omega}\frac{d\psi_{1}}{dx}=0
\end{displaymath}
Solving this equation, we will find $\psi_{1}$
\begin{equation}
\psi_{1}=(\frac{m\omega}{\pi\hbar})^{\frac{1}{4}}e^{-\frac{m\omega}{2\hbar}x^{2}}
\end{equation}
From (16) and (18),
\begin{equation}
\begin{array}{cc}
\psi_{n+1}=\sqrt{\frac{m \omega}{2n \hbar}}(x \psi_{n}-\frac{\hbar}{m \omega}\frac{d\psi_{n}}{dx}) & (n=1,2,\dots,s-1)
\end{array}
\end{equation}
Furthermore, we can get $\psi_{2}, \psi_{3}, \cdots, \psi_{s}$ in turn.
\begin{displaymath}
\psi_{2}=\sqrt{\frac{2m \omega}{\hbar}} x(\frac{m\omega}{\pi\hbar})^{\frac{1}{4}}e^{-\frac{m\omega}{2\hbar}x^{2}},\psi_{3}=\sqrt{2}(\frac{m \omega}{\hbar}x^{2}-\frac{1}{2})(\frac{m\omega}{\pi\hbar})^{\frac{1}{4}}e^{-\frac{m\omega}{2\hbar}x^{2}}, \cdots
\end{displaymath}
By means of mathematical induction, (1) will be proved from (19) and (20).
To summarize, it is not necessary to determine the matrix elements and the energy levels with invoking assistance from the bra-ket notation and its correlative definitions by Dirac.
\section{\bf Angular Momentum}
The angular momentum operator is a vector operator which can be written in terms of its vector components as:
\begin{displaymath}
\hat{\vec{M}}=\hat{M}_{x}\vec{i}+\hat{M}_{y}\vec{j}+\hat{M}_{z}\vec{k}
\end{displaymath}
where $\hat{\vec{M}}$ can be replaced as the orbital angular momentum operator $\hat{\vec{L}}$ and the total angular momentum operator $\hat{\vec{J}}$. The
components have the following commutation relations with each other:
\begin{displaymath}
\hat{M}_{y}=-\frac{i}{\hbar}[\hat{M}_{z},\hat{M}_{x}]
\end{displaymath}
\begin{displaymath}
\hat{M}_{x}=-\frac{i}{\hbar}[\hat{M}_{y},\hat{M}_{z}]
\end{displaymath}
A magnitude can be defined for the angular momentum operator:
\begin{displaymath}
\hat{M}^{2}=\hat{M}_{x}^{2}+\hat{M}_{y}^{2}+\hat{M}_{z}^{2}
\end{displaymath}
It commutes with the components of $\hat{\vec{M}}$
\begin{displaymath}
[\hat{M}^{2},\hat{M}_{x}]=[\hat{M}^{2},\hat{M}_{y}]=[\hat{M}^{2},\hat{M}_{z}]=0
\end{displaymath}
Let $M_{1}, M_{2},\cdots, M_{s}$ be the eigenvalue of the operator $\hat{M}_{z}$ and $Y_{1}, Y_{2},\cdots, Y_{s}$ be the orthonormalized
simultaneous eigenfunctions of the operators $\hat{M}_{z}$ and $\hat{M}^{2}$ respectively, then
\begin{displaymath}
\begin{array}{cc}
\hat{M}_{z}Y_{k}=M_{k}Y_{k} & (k=1,2,\cdots,s)
\end{array}
\end{displaymath}
Because $\hat{M}_{x}$ is a Hermitian operator, it is assumed that
\begin{displaymath}
\hat{M}_{x}\left[
\begin{array}{cccc}
Y_{1} & Y_{2} & \cdots & Y_{s} \\
\end{array}
\right]
=
\left[
\begin{array}{cccc}
Y_{1} & Y_{2} & \cdots & Y_{s} \\
\end{array}
\right]
\left[
\begin{array}{cccc}
M_{11} & M_{21}^{*} & \cdots & M_{s1}^{*} \\
M_{21} & M_{22} & \cdots & M_{s2}^{*} \\
\cdots & \cdots & \cdots & \cdots \\
M_{s1} & M_{s2} & \cdots & M_{ss} \\
\end{array}
\right]
\end{displaymath}
where $M_{11}, M_{22}, \cdots, M_{ss}$ are real numbers. Let $\hat{M}_{-}=\hat{M}_{x}-i\hat{M}_{y}$ and $\hat{M}_{+}=\hat{M}_{x}+i\hat{M}_{y} (c_{1}=c_{2}=-\frac{1}{\hbar})$ in terms of the theorem.
\begin{equation}
\begin{array}{cc}
M_{k}=(-\frac{s-1}{2}+k-1)\hbar & (k=1,2,\cdots,s)
\end{array}
\end{equation}
\begin{displaymath}
\hat{M}_{-}\left[
\begin{array}{ccccc}
Y_{1} & Y_{2} & \cdots & Y_{s-1} &  Y_{s} \\
\end{array}
\right]
=
\left[
\begin{array}{ccccccc}
Y_{1} & Y_{2} & \cdots & Y_{s-1} &  Y_{s} \\
\end{array}
\right]
\end{displaymath}
\begin{equation}
2\left[
\begin{array}{ccccc}
0 & M_{21}^{*} & \cdots & 0 & 0 \\
0 & 0 & \cdots & 0 & 0 \\
\cdots & \cdots & \cdots & \cdots & \cdots \\
0 & 0 & \cdots & 0 & M_{ss-1}^{*} \\
0 & 0 & \cdots & 0 & 0 \\
\end{array}
\right]
\end{equation}
\begin{displaymath}
\hat{M}_{+}\left[
\begin{array}{ccccc}
Y_{1} & Y_{2} & \cdots & Y_{s-1} &  Y_{s} \\
\end{array}
\right]
=
\left[
\begin{array}{ccccc}
Y_{1} & Y_{2} & \cdots & Y_{s-1} &  Y_{s} \\
\end{array}
\right]
\end{displaymath}
\begin{equation}
2\left[
\begin{array}{ccccc}
0 & 0 & \cdots & 0 & 0 \\
M_{21} & 0 & \cdots & 0 & 0 \\
\cdots & \cdots & \cdots & \cdots & \cdots \\
0 & 0 & \cdots & 0 & 0 \\
0 & 0 & \cdots & M_{ss-1} & 0 \\
\end{array}
\right]
\end{equation}
There is the commutation relation
\begin{displaymath}
2\hbar \hat{M}_{z}=[\hat{M}_{+},\hat{M}_{-}]
\end{displaymath}
So
\begin{displaymath}
2\hbar\left[
\begin{array}{ccccc}
M_{1}Y_{1} & M_{2}Y_{2} & \cdots & M_{s-1}Y_{s-1} &  M_{s}Y_{s} \\
\end{array}
\right]
\end{displaymath}
\begin{displaymath}
=2\hbar \hat{M}_{z}\left[\setlength\arraycolsep{0.3em}
\begin{array}{ccccc}
Y_{1} & Y_{2} & \cdots & Y_{s-1} &  Y_{s} \\
\end{array}
\right]
=[\hat{M}_{+},\hat{M}_{-}]\left[
\begin{array}{ccccc}
Y_{1} & Y_{2} & \cdots & Y_{s-1} &  Y_{s} \\
\end{array}
\right]
\end{displaymath}
\begin{displaymath}
=\!4\!\left[\setlength\arraycolsep{0.1em}
\begin{array}{ccccc}
\!-\!|M_{21}|^{2}Y_{1} & (|M_{21}|^{2}\!-\!|M_{32}|^{2})Y_{2} & \cdots & (|M_{s-1s-2}|^{2}\!-\!|M_{ss-1}|^{2})Y_{s-1} & |M_{ss-1}|^{2}Y_{s} \\
\end{array}
\right]
\end{displaymath}
Thus
\begin{displaymath}
\left
\{\begin{array}{c}
|M_{21}|^{2}=-\frac{\hbar}{2}M_{1},|M_{32}|^{2}=-\frac{\hbar}{2}(M_{1}+M_{2}),\cdots,|M_{ss-1}|^{2}=-\frac{\hbar}{2}\sum\limits_{p=1}^{s-1}M_{p} \\
|M_{ss-1}|^{2}=\frac{\hbar}{2}M_{s}
\end{array}\right.
\end{displaymath}
Furthermore, $M_{1}+M_{2}+\cdots+M_{s}=0$. Combining with (21),
\begin{displaymath}
\begin{array}{cccc}
M_{1}=-\frac{s-1}{2}\hbar, & M_{2}=-\frac{s-3}{2}\hbar, & \cdots, & M_{s}=\frac{s-1}{2}\hbar
\end{array}
\end{displaymath}
Therefore,
\begin{equation}
\begin{array}{cc}
\hat{M}_{z}Y_{k}=M_{k}Y_{k}=(-\frac{s-1}{2}+k-1)\hbar Y_{k} & (k=1,2,\cdots,s)
\end{array}
\end{equation}
\begin{equation}\setlength\arraycolsep{0.3em}
|M_{\!21}|^{2}\!=\!\frac{\hbar^{2}}{4}(\!s\!-\!1\!), |M_{\!32}|^{2}\!=\!\frac{\hbar^{2}}{4}2(\!s\!-\!2\!), \dots, |M_{\!ss-1}|^{2}\!=\!\frac{\hbar^{2}}{4}(\!s\!-\!1\!)[s\!-\!(\!s-\!1\!)]
\end{equation}
Let $|M_{10}|^{2}=0$, then we have in terms of (25)
\begin{displaymath}
\begin{array}{cc}
|M_{kk-1}|^{2}=\frac{\hbar^{2}}{4}(k-1)[s-(k-1)] & (k=1,2,\cdots,s)
\end{array}
\end{displaymath}
When we take positive real solutions, (22) can be written as
\begin{equation}
\begin{array}{cc}
\hat{M}_{-}Y_{k}=\hbar \sqrt{(k-1)[s-(k-1)]}Y_{k-1} & (k=1,2,\cdots,s)
\end{array}
\end{equation}
Let $|M_{s+1s}|^{2}=0$, then we have in terms of (25)
\begin{displaymath}
\begin{array}{cc}
|M_{k+1k}|^{2}=\frac{\hbar^{2}}{4}k(s-k) & (k=1,2,\cdots,s)
\end{array}
\end{displaymath}
When we take positive real solutions, (23) can be written as
\begin{equation}
\begin{array}{cc}
\hat{M}_{+}Y_{k}=\hbar \sqrt{k(s-k)}Y_{k+1} & (k=1,2,\cdots,s)
\end{array}
\end{equation}
We can express $\hat{M}_{x}$ and $\hat{M}_{y}$ in terms of $\hat{M}_{+}$ and $\hat{M}_{-}$. Thus
\begin{displaymath}
\hat{M}^{2}=\frac{1}{2}(\hat{M}_{+}\hat{M}_{-}+\hat{M}_{-}\hat{M}_{+})+\hat{M}_{z}^{2}
\end{displaymath}
Combining with (24) and (26)-(27),
\begin{equation}
\begin{array}{cc}
\hat{M}^{2}Y_{k}=\frac{\hbar^{2}}{4}(s^{2}-1)Y_{k} & (k=1,2,\cdots,s)
\end{array}
\end{equation}
\textbf{I: If s is an odd number}, the angular momentum operator $\hat{\vec{M}}$ is the orbital angular momentum operator $\hat{\vec{L}}$ which is defined as the cross product of the position vector $\vec{r}$ and the linear momentum operator $\hat{\vec{p}}$ of the particle.\\
\begin{displaymath}
\hat{\vec{L}}=\vec{r} \times \hat{\vec{p}}
\end{displaymath}
Let
\begin{displaymath}
\begin{array}{cc}
l=\frac{s-1}{2} & (s=1,3,5,\dots)
\end{array}
\end{displaymath}
\begin{displaymath}
\begin{array}{cc}
m=-l+k-1 & (k=1,2,\dots,s)
\end{array}
\end{displaymath}
If the wave functions $Y_{1},Y_{2},\cdots,Y_{s}$ is relabeled as $Y_{l-l},Y_{l1-l},\cdots,Y_{ll}$, then (24) and (26)-(28) can be written as
\begin{equation}
\begin{array}{cc}
\hat{L}_{z}Y_{lm}=m\hbar Y_{lm} & (l=0,1,\dots;m=-l,1-l,\cdots,l)
\end{array}
\end{equation}
\begin{equation}
\begin{array}{cc}
\hat{L}_{\!-}Y_{\!lm}\!=\!\hbar\sqrt{\!(l\!+\!m)(l\!+\!1\!-\!m)}Y_{\!lm\!-\!1} & (l=0,1,\dots;m=-l,1-l,\cdots,l)
\end{array}
\end{equation}
\begin{equation}
\begin{array}{cc}
\hat{L}_{\!+}Y_{\!lm}\!=\!\hbar\sqrt{\!(l\!+\!m\!+\!1)(l\!-\!m)}Y_{\!lm\!+\!1} & (l=0,1,\dots;m=-l,1-l,\cdots,l)
\end{array}
\end{equation}
\begin{equation}
\begin{array}{cc}
\hat{L}^{2}Y_{lm}=l(l+1)\hbar^{2}Y_{lm} & (l=0,1,\dots;m=-l,1-l,\cdots,l)
\end{array}
\end{equation}
In the spherical coordinate,
\begin{displaymath}
\left[\!
\begin{array}{ccc}
\!\vec{i} & \!\vec{j} & \!\vec{k}\!
\end{array}
\!\right]\!
\!\left[\!\setlength\arraycolsep{0.2em}
\begin{array}{c}
\hat{L}_{x}\\
\hat{L}_{y}\\
\hat{L}_{z}\\
\end{array}
\!\right]\!
\!=\!
\hat{\vec{L}}
\!=\!
\vec{r} \!\times\! \hat{\vec{p}}
\!=\!
\left[\!\setlength\arraycolsep{0.2em}
\begin{array}{ccc}
x & \!y & \!z\!
\end{array}
\right]\!
\!\left[\!\setlength\arraycolsep{0.2em}
\begin{array}{c}
\vec{i}\\
\vec{j}\\
\vec{k}\\
\end{array}
\!\right]\!
\!\times\!
\!\left[\setlength\arraycolsep{0.2em}
\begin{array}{ccc}
\!\vec{i}&\!\vec{j}&\!\vec{k}\!
\end{array}
\right]\!
\!\left[\!\setlength\arraycolsep{0.2em}
\begin{array}{c}
\hat{p}_{x}\\
\hat{p}_{y}\\
\hat{p}_{z}\\
\end{array}
\!\right]
\!=\!
\!\left[\!\setlength\arraycolsep{0.2em}
\begin{array}{ccc}
x & y & z
\end{array}
\!\right]\!
\!\left[\!\setlength\arraycolsep{0.2em}
\begin{array}{ccc}
\vec{0} & \vec{k} & -\!\vec{j}\\
-\!\vec{k} & \vec{0} & \vec{i}\\
\vec{j} & -\!\vec{i} & \vec{0}\\
\end{array}
\!\right]\!
\end{displaymath}
\begin{displaymath}
\!\left[\!\setlength\arraycolsep{0.2em}
\begin{array}{c}
-\!i\hbar\frac{\partial}{\partial x}\\
-\!i\hbar\frac{\partial}{\partial y}\\
-\!i\hbar\frac{\partial}{\partial z}\\
\end{array}
\!\right]\!
\!=\!
\!\left[\!\setlength\arraycolsep{0.2em}
\begin{array}{ccc}
z\vec{j}\!-\!y\vec{k} & x\vec{k}\!-\!z\vec{i} & y\vec{i}\!-\!x\vec{j}
\end{array}
\right]\!
\!(\!-\!i\hbar)\!\left[\!\setlength\arraycolsep{0.2em}
\begin{array}{c}
\frac{\partial}{\partial x}\\
\frac{\partial}{\partial y}\\
\frac{\partial}{\partial z}\\
\end{array}
\right]\!
\!=\!
\!(\!-\!i\hbar)\!\left[\!\setlength\arraycolsep{0.2em}
\begin{array}{ccc}
\vec{i} & \vec{j} & \vec{k}
\end{array}
\right]\!
\!\left[\!\setlength\arraycolsep{0.2em}
\begin{array}{ccc}
0 & -\!z & y\\
z & 0 & -\!x\\
-\!y & x & 0\\
\end{array}
\!\right]\!
\end{displaymath}
\begin{small}
	\begin{displaymath}
	\left[\!\setlength\arraycolsep{0.2em}
	\begin{array}{ccc}
	\frac{\partial r}{\partial x} & \frac{\partial \theta}{\partial x} & \frac{\partial \varphi}{\partial x}\\
	\frac{\partial r}{\partial y} & \frac{\partial \theta}{\partial y} & \frac{\partial \varphi}{\partial y}\\
	\frac{\partial r}{\partial z} & \frac{\partial \theta}{\partial z} & \frac{\partial \varphi}{\partial z}\\
	\end{array}
	\right]
	\left[\!\setlength\arraycolsep{0.2em}
	\begin{array}{c}
	\frac{\partial}{\partial r}\\
	\frac{\partial}{\partial \theta}\\
	\frac{\partial}{\partial \varphi}\\
	\end{array}
	\right]
	=
	(-i\hbar)\left[\!\setlength\arraycolsep{0.2em}
	\begin{array}{ccc}
	\vec{i} & \vec{j} & \vec{k}
	\end{array}
	\right]
	\left[\!\setlength\arraycolsep{0.2em}
	\begin{array}{ccc}
	0 & -r cos\theta & r sin\theta sin\varphi\\
	r cos\theta & 0 & -r sin\theta cos\varphi\\
	-r sin\theta sin\varphi & r sin\theta cos\varphi & 0\\
	\end{array}
	\right]
	\end{displaymath}
\end{small}
\begin{displaymath}
\left[\!\setlength\arraycolsep{0.2em}
\begin{array}{ccc}
sin\theta cos\varphi & \frac{cos\theta cos\varphi}{r} & -\frac{sin\varphi}{rsin\theta}\\
sin\theta sin\varphi & \frac{cos\theta sin\varphi}{r} & \frac{cos\varphi}{rsin\theta}\\
cos\theta & -\frac{sin\theta}{r} & 0\\
\end{array}
\right]\!
\!\left[\!\setlength\arraycolsep{0.2em}
\begin{array}{c}
\frac{\partial}{\partial r}\\
\frac{\partial}{\partial \theta}\\
\frac{\partial}{\partial \varphi}\\
\end{array}
\right]\!
\!=\!(\!-\!i\hbar)\!\left[\!\setlength\arraycolsep{0.2em}
\begin{array}{ccc}
\vec{i} & \vec{j} & \vec{k}
\end{array}
\right]\!
\!\left[\!\setlength\arraycolsep{0.2em}
\begin{array}{c}
sin\varphi \frac{\partial}{\partial \theta}\!+\!cot\theta cos\varphi \frac{\partial}{\partial \varphi}\\
-cos\varphi \frac{\partial}{\partial \theta}\!+\!cot\theta sin\varphi \frac{\partial}{\partial \varphi}\\
-\frac{\partial}{\partial \varphi}\\
\end{array}
\right]
\end{displaymath}
Therefore, the following expressions can be given by
\begin{equation}
\left \{  \begin{array}{c}
\hat{L}_{-}=\hbar e^{-i\varphi}(-\frac{\partial}{\partial\theta}+i\cot\theta\frac{\partial}{\partial\varphi})  \\
\hat{L}_{+}=\hbar e^{i\varphi}(\frac{\partial}{\partial\theta}+i\cot\theta\frac{\partial}{\partial\varphi}) \\
\hat{L}_{z}=-i\hbar\frac{\partial}{\partial\varphi}
\end{array}\right.
\end{equation}
with $\hat{L}_{-}=\hat{L}_{x}-i\hat{L}_{y},\hat{L}_{+}=\hat{L}_{x}+i\hat{L}_{y}$. (29)-(31) will form the overdetermined systems that there are more equations than unknowns. From (29), (30) and (33),
\begin{displaymath}
\left \{  \begin{array}{c}
\frac{\partial Y_{lm}}{\partial\varphi}=i m Y_{lm} \\
-\frac{\partial Y_{lm}}{\partial\theta}+i\cot\theta\frac{\partial Y_{lm}}{\partial\varphi}=\sqrt{(l+m)(l-m+1)}e^{i\varphi}Y_{lm-1}
\end{array}\right.
\end{displaymath}
When $m=-l$,
\begin{displaymath}
\begin{array}{c}
\left \{  \begin{array}{c}
\frac{\partial Y_{l-l}}{\partial\varphi}=-i l Y_{l-l} \\
\frac{\partial Y_{l-l}}{\partial\theta}=l\cot\theta Y_{l-l}
\end{array}\right.
\end{array}
\end{displaymath}
Solving this equation, we will find $Y_{l-l}$.
\begin{equation}
\begin{array}{cc}
Y_{l-l}=\sqrt{\frac{(2l+1)!}{4\pi}}\frac{sin^{l}\theta}{2^{l}l!}e^{-il\varphi} & (l=0,1,2,\cdots)
\end{array}
\end{equation}
Furthermore, from (31) and (33),
\begin{equation}
\begin{array}{cc}
Y_{lm+1}\!=\!\frac{1}{\sqrt{(l+m+1)(l-m)}}(\frac{\partial Y_{lm}}{\partial\theta}\!-\!m\cot\!\theta Y_{lm})e^{i\varphi} & (m=\!-l,1\!-\!l,\cdots,l\!-\!1)
\end{array}
\end{equation}
We can get $Y_{00},Y_{1-1},Y_{2-2},\cdots$ from (34) and other spherical harmonic functions are obtained from (35).\\
1) When $l$=0,
\begin{displaymath}
Y_{00}=\frac{1}{\sqrt{4\pi}}
\end{displaymath}
2) When $l$=1,
\begin{displaymath}
Y_{1-1}=\sqrt{\frac{3}{8\pi}}\sin\theta e^{-i\varphi}; Y_{10}=\sqrt{\frac{3}{4\pi}}\cos\theta, Y_{11}=-\sqrt{\frac{3}{8\pi}}\sin\theta e^{i\varphi}
\end{displaymath}
3) When $l$=2,
\begin{displaymath}
Y_{\!2-2}\!=\!\sqrt{\!\frac{15}{32\pi}}\sin^{2}\theta e^{-i2\varphi}; Y_{\!2-1}\!=\!\sqrt{\!\frac{15}{8\pi}}cos\theta \sin\theta e^{-i\varphi},Y_{\!20}\!=\!\sqrt{\!\frac{5}{16\pi}}(3cos^{2}\theta-1),
\end{displaymath}
\begin{displaymath}
Y_{21}=-\sqrt{\frac{15}{8\pi}}cos\theta \sin\theta e^{i\varphi},Y_{22}=\sqrt{\frac{15}{32\pi}}\sin^{2}\theta e^{i2\varphi}
\end{displaymath}
$\cdots,\cdots$\\
Let $x=cos\theta$, by means of mathematical induction, the following expression is proved from (34) and (35)
\begin{footnotesize}
	\begin{equation}
	\begin{array}{cc}
	Y_{lm}=(-1)^{m}\sqrt{\frac{(l-m)!}{(l+m)!}\frac{(2l+1)}{4\pi}}P^{m}_{l}(x) e^{im\varphi} & (l=0,1,\cdots;m=-l,1-l,\cdots,l-1)
	\end{array}
	\end{equation}
\end{footnotesize}
where $P^{m}_{l}(x)$ are the associated Legendre polynomials and $P^{m}_{l}(x)=\frac{(1-x^2)^{\frac{m}{2}}}{2^{l}l!}\\
\frac{d^{l+m}}{dx^{l+m}}(x^{2}-1)^{l}$. (36) are exactly the solutions to the second order differential equations
\begin{footnotesize}
	\begin{equation}
	\left \{ \begin{array}{l}
	-i\hbar\frac{\partial Y(\theta,\varphi)}{\partial\varphi}=\hat{L}_{z}Y(\theta,\varphi)=m\hbar Y(\theta,\varphi) \\
	-\hbar^{2}[\frac{1}{\sin\theta}\frac{\partial}{\partial\theta}(\sin\theta\frac{\partial}{\partial\theta})
	+\frac{1}{\sin^{2}\theta}\frac{\partial^{2}}{\partial\varphi^{2}}]Y(\theta,\varphi)=\hat{L}^{2}Y(\theta,\varphi)=l(l+1)\hbar^{2}Y(\theta,\varphi) \\
	\end{array}\right.
	\end{equation}
\end{footnotesize}
\textbf{II:  If s is an even number}, the angular momentum operator $\hat{\vec{M}}$ is the total angular momentum operator $\hat{\vec{J}}$. \\
Let
\begin{displaymath}
\begin{array}{cc}
j=\frac{s-1}{2} & (s=2,4,6,\dots)
\end{array}
\end{displaymath}
\begin{displaymath}
\begin{array}{cc}
m=-j+k-1 & (k=1,2,\dots,s)
\end{array}
\end{displaymath}
If the wave functions $Y_{1},Y_{2},\cdots,Y_{s}$ is relabeled as $Y_{j-j},Y_{j1-j},\cdots,Y_{jj}$, then we have similarly to the orbital angular momentum
\begin{displaymath}
\begin{array}{cc}
\hat{J}_{z}Y_{jm}=m\hbar Y_{jm} & (j=\frac{1}{2},\frac{3}{2},\dots;m=-j,1-j,\cdots,j)
\end{array}
\end{displaymath}
\begin{displaymath}
\begin{array}{cc}
\hat{J}_{-}Y_{jm}=\hbar\sqrt{(j+m)(j+1-m)}Y_{jm-1} & (j=\frac{1}{2},\frac{3}{2},\dots;m=-j,1-j,\cdots,j)
\end{array}
\end{displaymath}
\begin{displaymath}
\begin{array}{cc}
\hat{J}_{+}Y_{jm}=\hbar\sqrt{(j+m+1)(j-m)}Y_{jm+1} & (j=\frac{1}{2},\frac{3}{2},\dots;m=-j,1-j,\cdots,j)
\end{array}
\end{displaymath}
\begin{displaymath}
\begin{array}{cc}
\hat{J}^{2}Y_{jm}=j(j+1)\hbar^{2}Y_{lm} & (j=\frac{1}{2},\frac{3}{2},\dots;m=-j,1-j,\cdots,j)
\end{array}
\end{displaymath}
There are two schemes that unify the descriptions of matrix mechanics and wave mechanics.\\
Scheme I:\\
1) The wave functions are obtained by solving the second order differential equation. For example, we can get (36) from (37).\\
2) When the operators, which are treated as the signs of the derivatives, act on the wave functions, we can obtain the expressions that correspond to the system of differential equations as we can get (17) from (1) and (16). For example, we can also get (29)-(31) from (33) and (36).\\
3) The system of differential equations can be expanded to the matrix expressions as we can get (2) from (17). For example, we can also get the following expression from (30).\\
\begin{displaymath}
\hat{L}_{-} \begin{array}{ccccc}[Y_{2-2}&Y_{2-1}&Y_{20}&Y_{21}&Y_{22}]\end{array}
=\begin{array}{ccccc}[0&2\hbar Y_{2-2}&\sqrt{6}\hbar Y_{2-1}&\sqrt{6}\hbar Y_{20}&2\hbar Y_{21}]\end{array}
\end{displaymath}
\begin{displaymath}
=\begin{array}{ccccc}[Y_{2-2}&Y_{2-1}&Y_{20}&Y_{21}&Y_{22}]\end{array}
\hbar\left[
\begin{array}{ccccc}
0&2&0&0&0\\
0&0&\sqrt{6}&0&0\\
0&0&0&\sqrt{6}&0\\
0&0&0&0&2\\
0&0&0&0&0
\end{array}
\right]
\end{displaymath}
Scheme II:\\
1) To convert the operator relations into the matrix relations.\\
2) According to the relations between the matrices, the matrix elements will be determined.\\
3) The first order differential equations will be given to find the solution of equations.\\
Scheme II is adopted in this paper.\\

$\textbf{\emph{To be continue}}$
\end{document}